# Strain and thickness effects on magnetocrystalline anisotropy of CoFe(011) films

Eunsung Jekal*, Phuong Dao

*Department of Materials, ETH Zurich, 8093 Zurich, Switzerland*

We investigate MCA of CoFe(011) thin films as a function of strength of strain and film thickness has been studied. It is elucidated that perpendicular magnetocystalline anisotropy (MCA) energy ($E_{MCA}$) is getting stronger with compressed *xy*-plane lattice constant while in-plane MCA is become an easy-axis by tensile strain on *xy*-plane. The reason of the $E_{MCA}$ behaviors can be explained by features of electronic structures:

## I. INTRODUCTION

Materials in a form of thin films show very different properties from bulk because of the broken symmetry. Neel drew the attention of possible appearance of large magnetocrystalline anisotropy (MCA) of thin films. In addition, the systems with a strong perpendicular MCA energy ($E_{MCA}$) have potential to be used as high density

magnetic recording devices. [1–5]

For these applications, the corresponding materials have to be grown on different substrates. Due to the mismatch in lattice constants and thermal expansion coefficients between the magnetic material and the substrate, significant amounts of strain would be occurred, depending on the specific growth conditions and substrate materials.

In 1999, extremely large perpendicular magnetic anisotropy (PMA) effect was observed in CoFeB/MgO.

Since that time, a lot of experimental and theoretical research has been progressed to find origin of the large PMA of CoFeB/MgO. Then substrates, interface insertion layer, capping layers and other things are reported that they have an impact on PMA of CoFeB/MgO [6]. According to one experimental research, 100 nm CoFe/MgO(011) shows perpendicular $E_{MCA}$ about 0.75 erg/cm$^2$ while 100 nm CoFe/MgO(001) indicates 0.25 erg/cm$^2$.

Consequently, more than three times stronger $E_{MCA}$ is obtained by tailoring orientation of films.

In this study, we investigate magnetic properties of CoFe(011) thin films via density functional theory calculation. Furthermore, for consider the strain effect by the substrates, varied xy-plane lattice constant of [011] plane was considered.

## II. COMPUTATIONAL METHOD

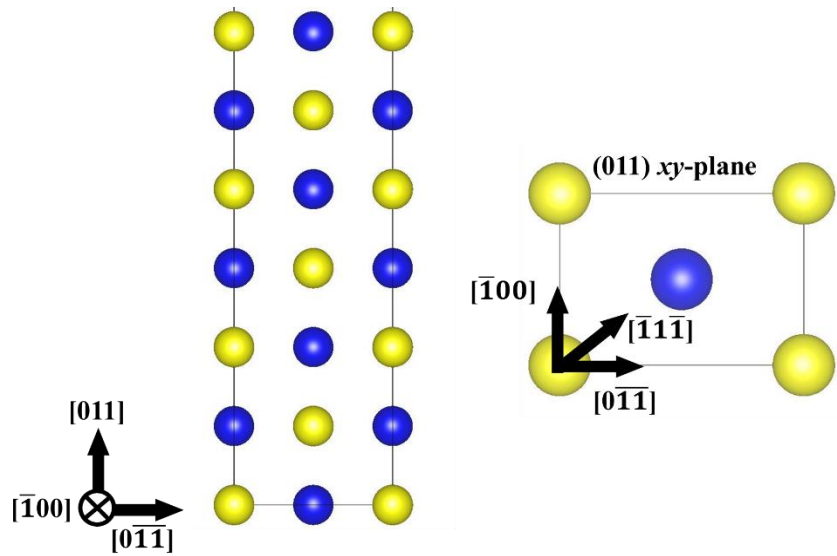

FIG. 1. The structure of the CoFe(011) film. Blue (yellow) spheres represent Co (Fe). Co and Fe atoms are placed in same layer with 1:1 ratio. Axes are presented in terms of coordinates of bcc structure.

First principles density functional theory (DFT) calculations [7, 8] were performed using the Vienna ab initio simulation package (VASP)[9]. We used the spin polarized generalized gradient approximation (GGA) [10] and projector augmented wave (PAW) potentials. 540 eV cut-off energy is used for wave function expansion. For our 2-D unit cell which is taken as rectangular with $1:\sqrt{2}$ ratio to describe the [011] plane of bcc structure, k-points sampling of a 16x12x1 mesh with Monkhorst-Pack generation scheme is generated. We adapted 20 Å spacing between adjacent layer to avoid artificial interaction in film calculations. Fig. 1 shows CoFe(011) film and its xy-plane. Blue and yellow spheres represent Co and Fe atoms, respectively. In case of CoFe(011) films, Co and Fe atoms are placed in same layer with 1:1 ratio. Axes are explicitly presented in terms of coordinates of bcc structure. We modeled CoFe(011) films with thickness of 2- to 7 ($2 \leq n \leq 7$)-mono-layers (MLs). To analyze strain effect, we adjusted xy-plane lattice constant from 2.68 Å to 3.00 Å

A while their atomic displacements along z-axis were fully relaxed until the Hellmann-Feynman forces were less than 0.1 meV/Å

To obtain $E_{MCA}$ of CoFe(011) films, we considered four directions of magnetization. As shown in Fig.1, $[0\bar{1}\bar{1}]$, $[\bar{1}00]$ and $[\bar{1}1\bar{1}]$ represent in-plane while $[011]$ represents out-of-plane.

Then we can obtain three kinds of $E_{MCA1}$: $E_{MCA1}=E_{0\bar{1}\bar{1}}-E_{011}$, $E_{MCA2}=E_{\bar{1}00}-E_{011}$, $E_{MCA3}=E_{\bar{1}0\bar{1}}-E_{011}$, where $E(0\bar{1}\bar{1})$, $E(100)$, $E(111)$ and $E(011)$ correspond to the total energies when the directions of the magnetization are along the where $(0\bar{1}\bar{1})$, $(100)$, $(111)$ and $(011)$, respectively.

Functions of $E_{MCA1}-E_{MCA2}=E_{(011)}-E_{(100)}$ and $E_{MCA2}-E_{MCA3}=E_{(100)}-E_{(111)}$ are energy to reverse magnetization on xy-plane. Convergence of the $E_{MCA}$ was checked with regard to cut off energy and number of k-point.

## III. RESULTS AND DISCUSSION

In Fig. 2, we present $E_{MCA1}$, $E_{MCA2}$ and $E_{MCA3}$ as a function of the xy-plane lattice constant with different film thicknesses from 2- to 7-MLs. Vertical dot lines represent 2.84 Å that lattice constant of bulk CoFe [11]. Larger and smaller values than 2.84 Å mean that we applied tensile and compressive strain on *xy*-plane, respectively.

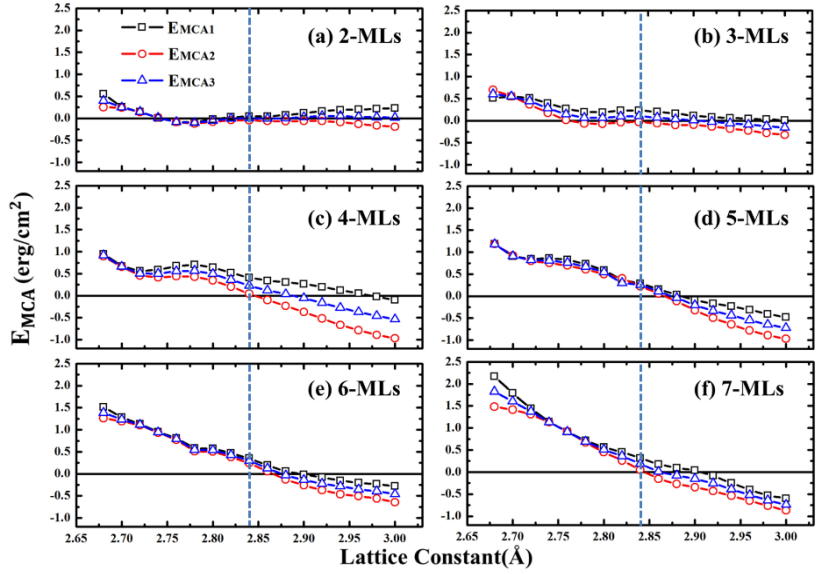

FIG.2 $E_{MCA}$ as a function of the lattice constant with various film thickness. Square (circle, triangle) correspond EMCA1($E_{MCA2}$,$E_{MCA3}$). Vertical dot lines represent lattice constant of bulk CoFe of 2.84 Å.

According to Fig. 2, n≥4 films show linear curve of $E_{MCA}$ variation that negative strain leads increased perpendicular $E_{MCA}$ but the easy-axis becomes in-plane with applying tensile strain. But when n=2 and 3, in-plane $E_{MCA}$ indicates around lattice constant of 2.76 Å and there is perpendicular $E_{MCA}$ while applied negative or positive strain since that lattice. One more important thing is that $E_{MCA}$ saturates in n≥5. According to Fig. 2(d), 2(e) and 2(f), films with a 5% decreased lattice constant of 2.68 Å has a saturated $E_{MCA}$ of about 1.5 erg/cm$^2$ and cases of 5% increased lattice constant of 3.00 Å how saturated $E_{MCA}$ f about -0.7 erg/cm$^2$

Fig. 3 presents magnetic moment of each layer from surface to center. We compared three cases which are 5% decreased, 5% increased and no strained lattice. In Fig. 3, magnetic moments of centered Fe atoms are reduced compared with surface atoms. In opposite to this, magnetic moments of Co are increased during Co atoms place far from surface. Total magnetic moments are increased (decreased) when apply positive (negative) strain.

For finding clues of the increased $E_{MCA}$ in strained films, we present density of states (DOS) of 2.68 Å and 3.00 Å lattice constant cases in n=2 and 7. n=2 shows different $E_{MCA}$ trend in comparison with thicker films, so we compare n=2 and 7. DOS of lattice constants of 2.68 Å and 3.00 Å are presented in Fig. 3, 4, respectively.

$E_{MCA}$ can be explained by perturbation theory [12].

According to perturbation theory, the spin orbit coupling (SOC) interaction between occupied and unoccupied states plays important role in determining $E_{MCA}$.

$E_{MCA}$ is expressed by

$$E_{MCA}^{\sigma\sigma'} \approx \xi^2 \sum_{o,u} \frac{|<o^\sigma|L_z|u^{\sigma'}>|^2 - |<o^\sigma|L_x|u^{\sigma'}>|^2}{\epsilon_{o,\sigma'} - \epsilon_{u,\sigma}},$$

where $\xi$ is the SOC strength, $\epsilon_{o,\sigma}$ ($\epsilon_{o,\sigma'}$) and $o^\sigma(u^\sigma)$ represent eigenvalues and eigenstates of occupied (unoccupied) for each spin state. When σσ' is ↑↑ or ↓↓, positive (negative) is determined by the SOC interaction between occupied and unoccupied states with the same (different by one) magnetic quantum number (m) through the $L_z$ ($L_x$) operator. However, in case of σσ=↑↓, positive (negative) contribution come from the $L_x$ ($L_z$) coupling [18].

In case of films with lattice constant of 2.68 Å which is a 5% reduced value compared to that of bulk, although both n=2 and 7 exhibit perpendicular $E_{MCA}$, $E_{MCA}$ of n=2 is about 0.252 erg/cm$^2$ while 7-MLs film exhibit much larger value of about 1.481 erg/cm$^2$. That causes are traceable to their DOS. According to Fig. 3, <yz ↓ |$L_x$|xy ↑ > couplings of Co of n=2 and 7 lead perpendicular anisotropy. However in Fig. 3(a) Co of n=2, occupied $z^2$ orbital is found nearby Fermi level. So <yz ↓ |$L_x$|$z^2$ ↓ > coupling abates perpendicular $E_{MCA}$ in n=2.

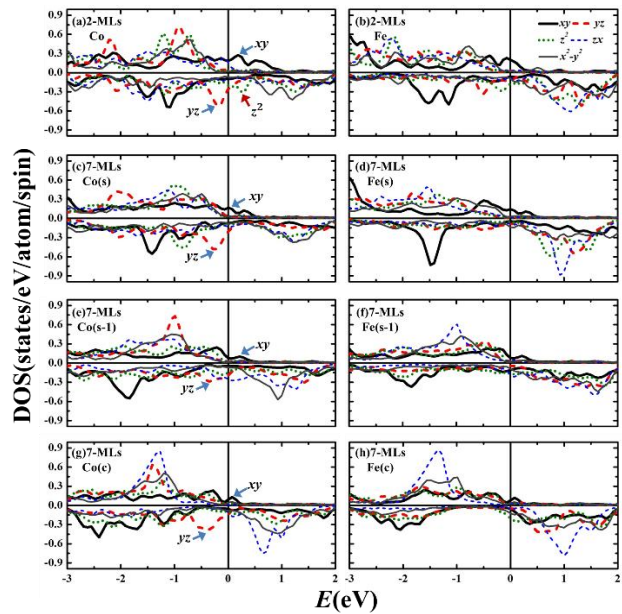

FIG. 4. DOS of n=2 and 7 with lattice constant of 2.62 Å. (a) Co and (b) Fe of n=2. (c) Co(S) and (d) Fe(S) of n=7. (e) Co(S-1) and (f) Fe(S-1) of n=7. (g) Co(C) and (h) Fe(C) of n=7. Thick solid, thick dash, dot, thin dash and thin solid line represent $xy$, $yz$, $z^2$, $zx$ and $x^2-y^2$ orbital, respectively.

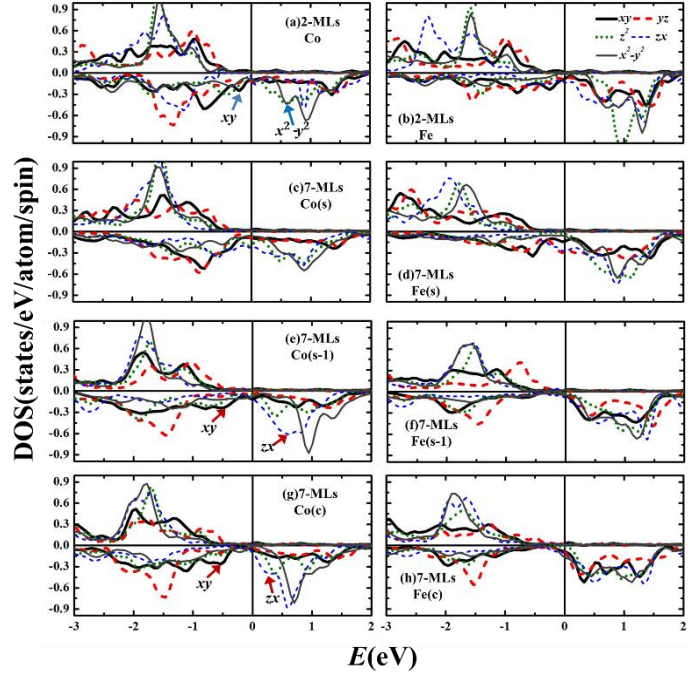

FIG. 5. DOS of n=2 and 7 with lattice constant of 2.62 Å. (a) Co and (b) Fe of n=2. (c) Co(S) and (d) Fe(S) of n=7. (e) Co(S-1) and (f) Fe(S-1) of n=7. (g) Co(C) and (h) Fe(C) of n=7.

## IV. CONCLUSION

In this study, MCA of CoFe(011) thin films as a function of strength of strain and film thickness has been studied. It is elucidated that perpendicular $E_{MCA}$ is getting stronger with compressed $xy$-plane lattice constant while in-plane MCA is become an easy-axis by tensile strain on xy-plane. The reason of the $E_{MCA}$ behaviors can be explained by features of electronic structures: $<yz\downarrow|L_x|xy\uparrow>$ couplings of Co atom play important role in determining perpendicular $E_{MCA}$ of tensile strained films, however,

$<xy\downarrow|L_x|zx\downarrow>$ coupling which contributes to in-plane MCA appears in compressive strained

films. Especially, the film with 10% tensile strain on the *xy*-plane exhibits perpendicular $E_{MCA}$ more than 1.00 erg/cm$^2$ for n≥5